\documentclass[sigconf]{acmart}

\AtBeginDocument{%
  }

\setcopyright{acmlicensed}
\copyrightyear{2024}
\acmYear{2024}
\setcopyright{acmlicensed}\acmConference[ASE '24]{39th IEEE/ACM International Conference on Automated Software Engineering }{October 27-November 1, 2024}{Sacramento, CA, USA}
\acmBooktitle{39th IEEE/ACM International Conference on Automated Software Engineering (ASE '24), October 27-November 1, 2024, Sacramento, CA, USA}
\acmDOI{10.1145/3691620.3695000}
\acmISBN{979-8-4007-1248-7/24/10}

\usepackage{graphicx}
\usepackage{colortbl}
\usepackage{multirow}
\usepackage{pifont}
\usepackage{subcaption}
\usepackage{adjustbox}
\usepackage{algorithm}
\usepackage{algpseudocode}
\usepackage{fixltx2e}
\usepackage{enumitem}

\newcommand{\find}[1]{
	\setlength{\fboxrule}{1pt}
	\begin{center}\
		\noindent\fcolorbox{black}{gray!10}{
		
		\begin{minipage}{.92\linewidth}
			#1
		\end{minipage}
	}
	\end{center}
	\smallskip
}

\begin{document}

%%
%% The "title" command has an optional parameter,
%% allowing the author to define a "short title" to be used in page headers.
\title[short title]{Sifting through the Chaff: On Utilizing Execution Feedback for Ranking the Generated Code Candidates}
\title[On Utilizing Execution Feedback for Ranking the Generated Code Candidates]{Sifting through the Chaff: On Utilizing Execution Feedback for Ranking the Generated Code Candidates}
% \title{Utilizing Execution Feedback for Generated Code Candidates Ranking}

%%
%% The "author" command and its associated commands are used to define
%% the authors and their affiliations.
%% Of note is the shared affiliation of the first two authors, and the
%% "authornote" and "authornotemark" commands
%% used to denote shared contribution to the research.
\author{Zhihong Sun}
\affiliation{%
  \institution{Shandong Normal University}
  \city{Jinan}
  \country{China}
}
\email{2022021002@stu.sdnu.edu.cn}

\author{Yao Wan}
\affiliation{%
  \institution{Huazhong University of Science and Technology}
  \city{Wuhan}
  \country{China}}
\email{wanyao@hust.edu.cn}

\author{Jia Li}
\affiliation{%
  \institution{Key Lab of HCST (PKU), MOE; SCS}
  \city{Beijing}
  \country{China}}
\email{lijiaa@pku.edu.cn}

\author{Hongyu Zhang}
\affiliation{%
  \institution{Chongqing University}
  \city{Chongqing}
  \country{China}}
\email{hyzhang@cqu.edu.cn}

\author{Zhi Jin}
% \authornote{Zhi Jin and Chen Lyu are the corresponding authors.}
\affiliation{%
  \institution{Key Lab of HCST (PKU), MOE; SCS}
  \city{Beijing}
  \country{China}}
\email{zhijin@pku.edu.cn}

\author{Ge Li}
\affiliation{%
  \institution{Key Lab of HCST (PKU), MOE; SCS}
  \city{Beijing}
  \country{China}}
\email{lige@pku.edu.cn}

\author{Chen Lyu}
% \authornotemark[1]
\authornote{Corresponding author.}
\affiliation{%
  \institution{Shandong Normal University}
  \city{Jinan}
  \country{China}
}
\email{lvchen@sdnu.edu.cn}

%%
%% By default, the full list of authors will be used in the page
%% headers. Often, this list is too long, and will overlap
%% other information printed in the page headers. This command allows
%% the author to define a more concise list
%% of authors' names for this purpose.
\renewcommand{\shortauthors}{Sun et al.}

\begin{abstract}
  \textit{Large Language Models} (LLMs), such as GPT-4, StarCoder, and Code Llama, are transforming the way developers approach programming by automatically generating code based on given contexts, such as natural language descriptions or incomplete surrounding code. Despite advancements, generating syntactically and semantically correct code remains challenging, especially for complex programming tasks. 
  % \textcolor{red}{
  % Typically, individuals generate multiple candidate solutions using LLMs to increase the likelihood of producing correct code.} 
  Existing approaches typically generate multiple candidate solutions using LLMs to increase the likelihood of producing correct code.
  However, selecting the correct code from these candidates — a process known as code ranking — remains a major challenge. Current research on code ranking can be categorized into execution-based and non-execution-based methods. Execution-based methods, although effective, encounter notable limitations, such as scarcity of quality unit tests and security risks. Non-execution-based methods like CodeRanker, which rely solely on classification labels to train a code ranker, struggle to capture subtle errors and provide detailed error insights. Recognizing the strengths and limitations of both approaches, we propose a new method that integrates the advantages of execution-based and non-execution-based techniques. The key insight of our work is that an effective code ranker is expected to truly comprehend the underlying causes of erroneous code, as relying solely on classification labels is insufficient. Inspired by this, this paper puts forward RankEF, an innovative approach for code ranking that leverages execution feedback. RankEF employs multi-task learning to integrate code classification with execution feedback generation. This approach enables the model to understand the reasons behind incorrect code, distinguishing between correct and incorrect solutions without the need to execute the code during the ranking phase. Experiments on three code generation benchmarks—APPS, MBPP, and HumanEval—demonstrate that RankEF significantly outperforms the state-of-the-art CodeRanker, achieving relative improvements of +30.97\%, +31.43\%, and +19.51\% in Pass@1, Pass@2, and Pass@5 on APPS test, respectively.
\end{abstract}

%
% The code below is generated by the tool at http://dl.acm.org/ccs.cfm.
% Please copy and paste the code instead of the example below.
%


%%
%% Keywords. The author(s) should pick words that accurately describe
%% the work being presented. Separate the keywords with commas.
%% A "teaser" image appears between the author and affiliation
%% information and the body of the document, and typically spans the
%% page.

% \received{20 February 2007}
% \received[revised]{12 March 2009}
% \received[accepted]{5 June 2009}

%%
%% This command processes the author and affiliation and title
%% information and builds the first part of the formatted document.
\maketitle

\section{Introduction}
% Recently, we have witnessed a significant surge of \textit{Large Language Models (LLMs)} for code, including GPT-4~\cite{GPT-4}, StarCoder~\cite{li2023starcoder}, and Code Llama~\cite{codellama}, which revolutionize the field of code generation. Although LLMs have demonstrated strong capabilities in code generation, 
% generating programs without syntax and runtime errors 
% remains a challenging endeavor, as it necessitates a comprehensive understanding of the programming task and the code generation process~\cite{inala2022fault}.
Recently, we have witnessed a significant surge in the development of \textit{Large Language Models (LLMs)} for code, including GPT-4~\cite{GPT-4}, StarCoder~\cite{li2023starcoder}, and Code Llama~\cite{codellama}, which have revolutionized the field of code generation. Despite their strong capabilities, generating programs without syntax and runtime errors remains a challenging endeavor, as it requires a comprehensive understanding of both the programming task and the code generation process~\cite{li2022competition, hendrycks2021measuring}.

% their performance still falls short in tackling certain complex programming problems~\cite{li2022competition}. However, retraining an LLM requires substantial computational resources. Therefore, exploring some post-processing methods to improve the code generation capabilities of LLMs without retraining a new model becomes very appealing.

%in the face of complex programming problems 
% It has been observed that although LLMs perform poorly when handling complex programming problems, they are often able to produce correct code when generating a large number of candidate codes
% Although LLMs perform poorly on complex programming problems, they are often able to produce correct code when generating a large number of candidate solutions for the same problem. This phenomenon has intrigued researchers, prompting them to conduct a series of studies on how to select the correct code from among the many candidate codes, known as code ranking. Currently, research on code ranking can be divided into two main categories based on \textit{whether or not code execution is required during the ranking process}: execution-based code ranking and non-execution-based code ranking.
Although LLMs often struggle with complex programming problems, they can frequently produce correct code when generating multiple candidate solutions. This phenomenon has intrigued researchers, prompting a series of studies focused on selecting the correct code from multiple candidates, a process referred to as code ranking. 
% Current research on code ranking can be divided into two main categories based on whether code execution is required during the ranking process: execution-based code ranking and non-execution-based code ranking.
Current research on code ranking can be divided into two main approaches: execution-based, which requires running the code during ranking, and non-execution-based, which does not.

Execution-based code ranking methods, which involve sampling a large number of code candidates and executing a few unit tests to filter out those that pass, have notably improved the code generation capabilities of LLMs~\cite{humaneval, austin2021program, li2022competition}. 
For example, 
MBR-EXEC~\cite{MBR-EXEC} introduced a method that executes a small number of test cases and filters code based on the Minimum Bayes Risk (MBR) of the execution results. Subsequently, CodeT~\cite{chen2022codet} addressed the challenge of obtaining unit tests by proposing the generation of unit tests using LLMs, and then filtering the code with these generated tests. 
% While these execution-based methods can accurately filter out correct code, their applicability in practice is constrained as obtaining quality unit test cases is time-consuming~\cite{siddiq2023exploring, schafer2023empirical, lukasczyk2023empirical}, and executing code generated by LLMs without proper security guarantees poses significant security risks~\cite{siddiq2023generate, toth2024llms}. 
While execution-based methods can effectively filter correct code, their practical applicability is limited, as generating high-quality unit test cases is time-consuming~\cite{siddiq2023exploring, schafer2023empirical, lukasczyk2023empirical}, and executing code generated by LLMs without robust security guarantees introduces significant security risks~\cite{siddiq2023generate, toth2024llms}.
Consequently, researchers have shifted focus to non-execution-based code ranking methods. The most notable among these is CodeRanker~\cite{inala2022fault}, which uses the execution results (correct or incorrect) of code as supervised signals to train a code classifier. However, relying solely on classification labels does not allow the code classifier to learn the specific reasons for code errors, thereby hindering its ability to accurately distinguish between correct and incorrect code.

We identify several key issues with CodeRanker:
\textbf{\ding{182} Inadequate detection of subtle errors.} Subtle error types, such as syntax errors, index errors, and undefined variable references, cannot be accurately captured by simple classification labels. Although CodeRanker attempts to refine classification labels into more granular error types, this approach has not shown significant improvement and has sometimes even worsened performance. This indicates that merely increasing the number of categories does not effectively enhance the model's capability.
\textbf{\ding{183} Insufficient error context.} Utilizing code execution results as classification labels inherently limits the information provided, indicating only whether the code is correct or incorrect. This approach does not reveal detailed reasons for errors, such as specific error messages or error locations, thereby lacking the granularity necessary for a comprehensive understanding of code faults.
\textbf{\ding{184} Weak generalization.} Models trained solely on classification labels are highly sensitive to label distribution, leading to weak generalization ability, as demonstrated in numerous studies~\cite{larson2021exploring, deng2022fifa}.

% The issues highlighted above 
These issues motivate us to develop a method that enables the code ranker to comprehensively understand the reasons behind code errors, thereby allowing it to accurately differentiate between correct and incorrect code. 
We find that through executing code, we obtain not only the results but also valuable feedback, including the error's location in the code, error messages, and detailed error types. Such feedback can help programmers identify the root causes of code errors. However, current research on non-execution-based code ranking has not yet effectively leveraged this information.

%Based on the above idea, in this paper, we propose a novel approach termed RankEF, which utilizes the \underline {\textbf{E}}xecution \underline{\textbf{F}}eedback for code \underline{\textbf{Rank}}ing in code generation. To integrate execution feedback into non-execution-based code ranking methods, we face the following three key challenges:

In this paper, we propose a novel approach termed RankEF, which utilizes the \underline {\textbf{E}}xecution \underline{\textbf{F}}eedback for code \underline{\textbf{Rank}}ing in code generation. To overcome the limitations of execution-based methods, RankEF does not execute code during the ranking phase (i.e., the inference phase). 
% This requirement poses three major challenges for effectively utilizing execution feedback in RankEF.
This introduces three major challenges in effectively leveraging execution feedback within RankEF.

\noindent
\textbf{Challenge 1: Overcoming the dependence of execution feedback during ranking}. 
%\hy{is it Overcoming the dependence of execution feedback during ranking}\lc{done.}
%\hy{this challenge is not clear, do you mean: 'Overcoming the reliance on execution feedback'?, also, is it overcoming or reducing? did you use code execution feedback?} \lc{Added at the top of the paragraph is an explanation of why rely on execution feedback during the ranking phase}. 
% An intuitive approach to leveraging execution feedback is to use it as an input to the model, guiding the model to understand the causes of code errors. 
An intuitive approach to leveraging execution feedback is to use it as input for the model, guiding it to better understand the underlying causes of code errors.
However, this approach still relies on execution feedback during the ranking phase, leading to the same potential limitations as execution-based code ranking methods. \textbf{To address this challenge}, our RankEF employs a multi-task learning approach. By using execution feedback and code execution results as  supervised signals during the training phase, we train the code ranker through a combination of code classification and execution feedback generation tasks. This approach enables the code ranker to distinguish between correct and incorrect code while also understand the root causes of errors in erroneous code. Since execution feedback is only needed during the training phase, it eliminates the dependence on execution feedback during the actual ranking phase.
% \hy{the descriptions here are not clear}.
% \lc{done.}

%\textbf{Challenge\#2: Disorganized execution feedback}. 
% \textbf{Challenge\#2: Inconsistent and noisy execution feedback}. 
% Although code execution feedback contains a wealth of useful information, the format of feedback from different code executions is often inconsistent, and some errors may stem from external packages on which the code depends, unrelated to the current code. \textbf{To address this challenge}, we designed different templates based on various code execution results and utilized regular expressions to extract useful information, integrating it into the templates. The templated execution feedback ensures consistency in training data, reduces data noise, and improves data quality, enabling the model to more accurately understand and learn characteristics. This approach has been validated by several studies~\cite{garcia2015data, li2017feature}.

\noindent
\textbf{Challenge 2: Inconsistent and noisy execution feedback}. 
% Code execution feedback, while rich in valuable information, often suffers from inconsistency in format and may include errors arising from external dependencies rather than the code itself. \textbf{To address this challenge}, we designed specific templates for various types of code execution results and employed regular expressions to extract relevant information, integrating it into these templates
Code execution feedback, while rich in valuable information, often suffers from inconsistent formats and may include errors arising from external dependencies rather than the code itself. \textbf{To address this challenge}, we categorize all code execution results into three classes: correct, execution error (i.e., compilation and runtime issues), and intent error (i.e., input-output mismatch error). We develop three templates for various types of code execution results and used regular expressions to extract and integrate relevant information. 
This templated execution feedback ensures consistency in the training data, reduces noise, and improves data quality, enabling the model to more accurately understand and learn from code. This approach has been validated by several studies~\cite{garcia2015data, li2017feature}.

%\textbf{Challenge\#3: Effective use of execution feedback}. 
% \textbf{Challenge\#3: Balancing multi-task learning}. 
% Since we employ multi-task learning, which is known to have its pros and cons, improper use of this method may lead to conflicts between the two tasks, potentially severely impacting the performance of the code ranker. \textbf{To address this challenge}, we designed three different multi-task learning methods to explore the optimal strategy that maximizes the utility of execution feedback. Additionally, we conducted detailed experimental analysis on the weighting of the two tasks to ensure a reasonable balance and overall performance optimization.

\noindent
\textbf{Challenge 3: Balancing multi-task learning}. 
Employing multi-task learning, while beneficial, can lead to conflicts between tasks if not properly managed, potentially affecting the performance of ranker. \textbf{To address this challenge}, we design three multi-task learning strategies (i.e., hard parameter sharing, soft parameter sharing, and intermediate fine-tuning) to identify the optimal approach for maximizing the utility of execution feedback. Additionally, we perform a detailed experimental analysis on the task weighting to ensure a balanced and optimized overall performance.

To validate the effectiveness of RankEF, we perform experiments on three code generation benchmarks, namely APPS~\cite{hendrycks2021measuring}, MBPP~\cite{austin2021program}, and
HumanEval~\cite{humaneval}.
Experimental results show that RankEF outperforms the state-of-the-art CodeRanker, achieving relative improvements of +30.97\%, +31.43\%, and +19.51\% for Pass@1, Pass@2, and Pass@5 on APPS test, respectively. Additionally, RankEF demonstrates superiority compared to other non-execution methods.

In summary, the key contributions of this paper are three-fold.

\setlist[itemize]{left=0pt}
\begin{itemize}
\item We are the first to leverage both the supervised signals offered by classification labels and the feedback derived from code execution to rank the generated code candidates.

\item To effectively utilize executive feedback, we design a unified multi-learning framework, in which three distinct strategies (i.e., hard parameter sharing, soft parameter sharing, and intermediate fine-tuning) are explored.

\item We conduct extensive experiments utilizing nine LLMs across three widely recognized code generation benchmarks (i.e., APPS, MBPP, and HumanEval). The results show that our proposed approach significantly outperforms existing baselines.
\end{itemize}

\section{Motivation}
\subsection{A Motivating Example}

\begin{figure}[htbp]
    \centering
    \includegraphics[width=\linewidth]{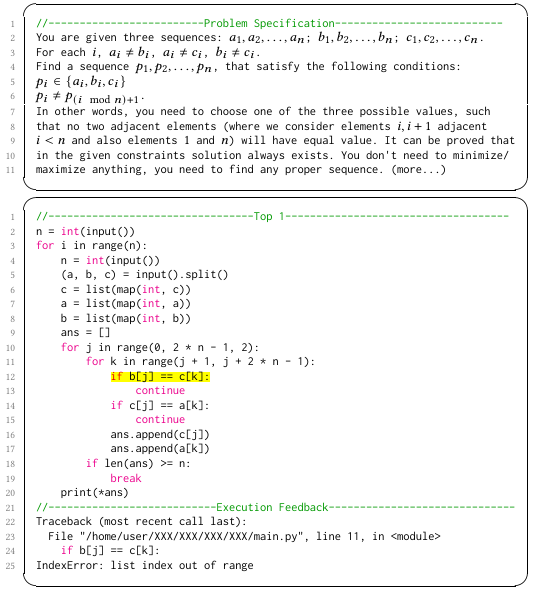}
    \caption{An example after being ranked by CodeRanker.}
    \vspace{-1mm}
    \Description{An example after being ranked by CodeRanker.}
    \label{fig: motivation}
    \vspace{-1mm}
\end{figure}

Figure~\ref{fig: motivation} presents a candidate code snippet generated by CodeT5 based on the problem specification. CodeRanker places this code at the top of its list. However, this code fails to meet the user's requirements. At a glance, even experienced programmers might struggle to identify the flaw in the code without executing it to obtain feedback. The challenge lies in CodeRanker's current training methodology, which relies solely on execution results (whether the code is correct or incorrect) as the supervisory signal. This approach does not allow CodeRanker to grasp the underlying reasons why a particular code is erroneous, thereby impairing its ability to accurately distinguish between correct and incorrect code. 

\begin{figure*}
    \centering
    \includegraphics[width=\linewidth]{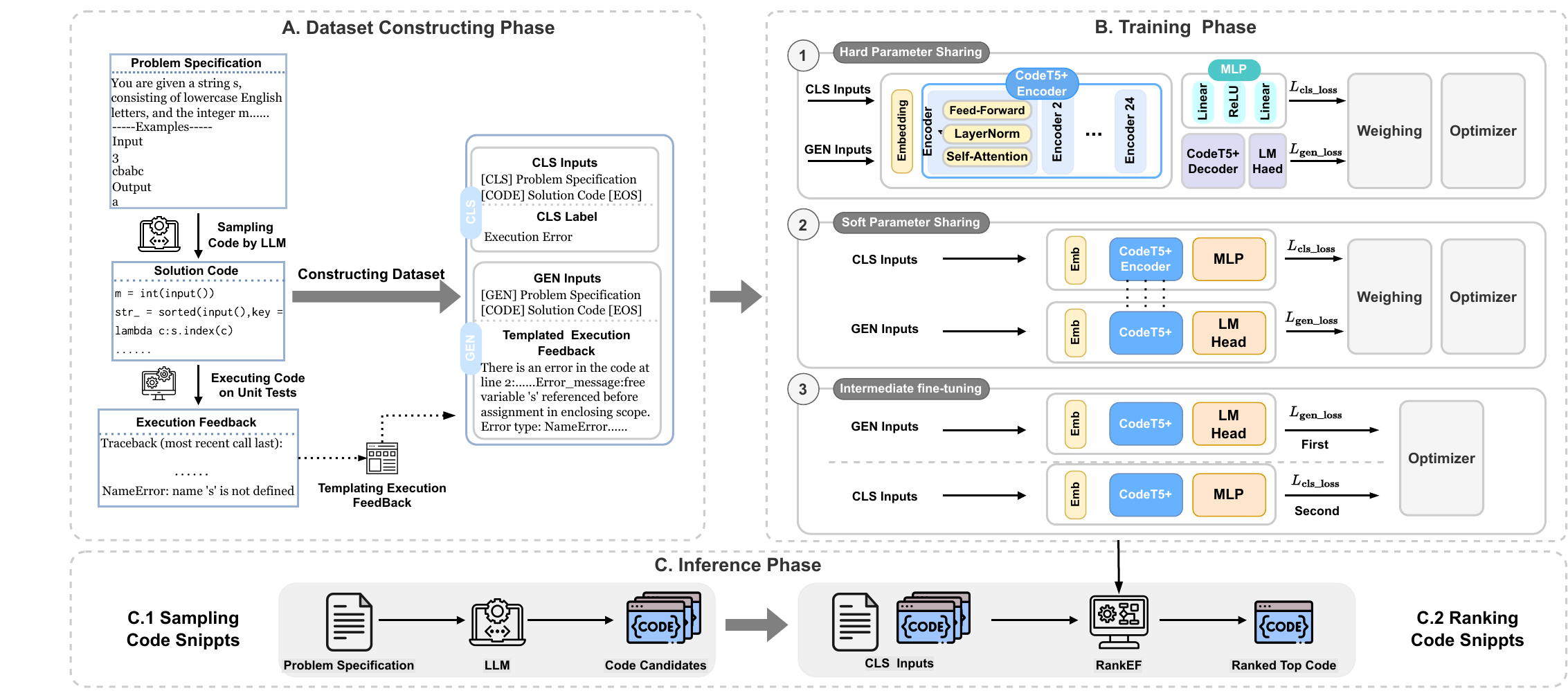}
    \caption{The architecture of RankEF. Phase \textbf{A.} Dataset construction tailored for RankEF, where CLS Inputs and GEN Inputs represent the inputs for the classification task and the inputs for the execution feedback generation task, respectively. Phase \textbf{B.} Diverse multi-task training strategies for RankEF: \ding{172} Hard Parameter Sharing. \ding{173} Soft Parameter Sharing. \ding{174} Intermediate Fine-Tuning. Phase \textbf{C.} Ranking process with RankEF.}
    \Description{The architecture of RankEF.}
    \label{fig: model}
    \vspace{-4mm}
\end{figure*}

Executing the code snippet allows us to obtain execution feedback from the compiler. From this example, we can obtain a lot of useful information from the execution feedback. We observe an \textit{IndexError} occurring at line 11: \texttt{if b[j] == c[k]:}. This error arises because of an out-of-range index, revealing a fundamental flaw in the code logic. Understanding the nature of such errors is crucial, as execution feedback provides valuable insights into why the code fails. This information plays an important role in helping the model understand the causes of code errors and improve ranking capabilities. These observations motivate us to explore methods that effectively leverage execution feedback for code ranking. By incorporating detailed execution feedback into our training process, we aim to enable our model to better comprehend and differentiate between correct and incorrect code, ultimately improving code ranking accuracy.

\subsection{Key Idea}
% The key idea of our approach revolves around the notion that execution feedback, which offers context regarding the background or circumstances in which an error occurred, carries valuable insights that aid the ranker in gaining a deeper understanding of the reasons and underlying issues behind the errors. Therefore, it is imperative to concurrently account for both the supervisory signals provided by classification labels and the feedback derived from code execution in the context of code ranking. To address this, we formulate a unified multi-task learning framework that accommodates these two supervisory signals. This framework is designed to facilitate RankEF in its dual objectives of generating feedback and classifying code, ultimately enabling it to discern between correct and erroneous code while delving deeper into the underlying sources of errors.

%We believe that an ideal ranker should simultaneously consider the supervisory signals provided by classification labels and code execution feedback. 
Based on our analysis, we assert that an effective ranker is expected to integrate two crucial sources of information: the supervisory signals derived from classification labels and the insights gained from code execution feedback. This integration ensures a comprehensive evaluation that captures both the static code features and dynamic execution behavior, leading to more accurate and reliable code ranking outcomes.
%This is because execution feedback offers valuable insights into the context and causes of errors, aiding the ranker in gaining a deeper understanding of the underlying reasons behind the errors.
To equip the ranker with these capabilities, the key idea of our approach is to develop a unified multi-task learning framework that integrates both classification labels and code execution feedback. From the perspective of the code classification task, integrating classification labels' supervisory signals enables the ranker to possess basic classification capabilities, effectively distinguishing between correct and erroneous code. From the perspective of the execution feedback generation task, incorporating code execution feedback allows the ranker to deeply understand and analyze the root causes of errors, thereby enhancing the accuracy and effectiveness of the ranking.

\section{Approach}

We now introduce the detailed formulation and training procedures of RankEF. The detailed process is illustrated in Figure ~\ref{fig: model}.

% \subsection{Constructing datasets for RankEF}
\subsection{Problem Definition}

Given a set of code candidates $S = \{S_1, \dots, S_n\}$, produced by a code generation model, the objective of RankEF is to rank these code candidates so that the accurate code comes first. 
For each natural language description $N$ and corresponding code candidate $S_i$, RankEF processes the input in a specific format. The format is $\texttt{[CLS]}\texttt{[QUERY]},n_1,\dots,n_{|N|},\texttt{[CODE]},s_{i_{1}},\dots, s_{|S_{i}|},\texttt{[EOS]}$, where $N$ and $S_i$ are concatenated by special separators. These separators include $\texttt{[CLS]}$, which is similar to BERT~\cite{devlin2018bert} and used to indicate the full information of the input, signifying that this is a classification task, and lowercase letters to indicate the tokens of $N$ and $S_i$.
Using this, RankEF calculates a score $p_i$ indicating the correctness of $S_i$, and then ranks the candidates based on these scores, with a goal to position the correct code higher in the list. 

\begin{figure}
    \centering
    \includegraphics[width=\linewidth]{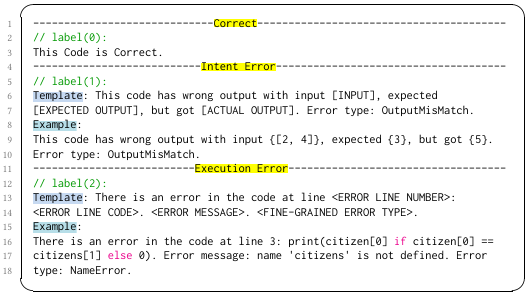}
    \caption{An example of templated execution feedback.}
    \vspace{-1mm}
    \Description{An example of templated execution feedback.}
    \label{fig: template}
    \vspace{-3mm}
\end{figure}
\subsection{Dataset Construction}

To train RankEF, we first sample code from the model and gather execution feedback, forming a crucial part of our dataset, as shown in Phase \textbf{A} of Figure~\ref{fig: model}. Using various base code models, we sample code from the APPS training set. Each problem in APPS yields a sample of 100 code snippets, which are then de-duplicated and executed for feedback. The execution outcomes categorized the code into three groups: $\{$$Correct$, $Intent\,Error$, $Execution\, Error$$\}$. 

Since the execution feedback of the code contains some messy and useless information, we keep only the information that is useful for understanding the code, including fine-grained error types, error line numbers, error line code, and mismatched inputs and outputs, etc. We combined this helpful information following a designed template as illustrated in Figure ~\ref{fig: template}    . For correct code, we define its execution feedback as “This code is correct”. For code with the intent error, we combine the relevant inputs, expected outputs, and actual outputs of the code as execution feedback. For code that executes incorrectly, we combine fine-grained error types, error message, error line code, etc.

This process culminates in a quadruple $(N, S, E, F)$, where $E$ is the classification label and $F$ represents templated feedback. Given the variation in error types across models, it is vital to craft a specific code ranker for each model.

\subsection{Mutil-Task Training for RankEF}
CoderRanker strictly treats code ranking as a classification task, and its training objective can be defined as:

\begin{equation}
\mathcal{L}_{cls\_loss}=-\sum_i^n E_i \log P\left(\hat{E}_i \mid N, S\right)
\end{equation}
However, solely relying on classification tasks results in a ranker with limited discernment. To overcome this, we adopt a multi-task training strategy, supplementing the classification with a generation task. This addition helps RankEF discover the root causes of code errors using execution feedback. For this, we utilize a model input format, $GEN\_S = \texttt{[GEN]}\texttt{[QUERY]},n_1,\dots,n_{|N|},\texttt{[CODE]},s_{i_{1}},\dots,s_{|S_{i}|},$ $\texttt{[EOS]}$, where RankEF predicts the nature of the code errors. The objective for this generation task is defined as:
\begin{equation}
\mathcal{L}_{gen\_loss }=-\sum \log P\left( F \mid N, S\right)
\end{equation}
% \textcolor{red}{where F denotes xxxx? N denotes xxx, S denotes xxx?}
The generation task bolsters the classification task by deepening the model's understanding of why code failed, improving its ability to distinguish correct from incorrect code. Additionally, we investigate three distinct multi-task training methodologies, as depicted in Phase \textbf{B} of Figure~\ref{fig: model}.
\noindentparagraph{\textbf{\textup{Hard Parameter Sharing.}}}
In hard parameter sharing, our model's encoder component shares parameters for both tasks. The code ranking model integrates the classification and generation tasks, optimizing both the classification loss (`cls\_loss') and generation loss (`gen\_loss') within a unified model framework. The learning objective for hard parameter sharing can be formally described as:
\begin{equation}
Loss=(1-\lambda)\mathcal{L}_{cls\_loss}+\lambda \mathcal{L}_{gen\_loss}
\end{equation}

% \paragraph{Soft Parameter Sharing.}
\noindentparagraph{\textbf{\textup{Soft Parameter Sharing.}}}
In soft parameter sharing, analogous to hard parameter sharing, there are two learning objectives employed to train the code ranking model. The distinction lies in the fact that soft parameter sharing entails training separate models for each task, resulting in one classification model and one generation model. To facilitate knowledge sharing between these models, we incorporate a shared loss function utilizing the Euclidean norm~\cite{golub2013matrix}:
% \begin{equation}
%     \begin{adjustbox}{max width=0.85\linewidth}
%         $\mathcal{L}_{\mathrm{sharing\ }}\left(\omega_1,\omega_2\right)=\sqrt{\sum_{i=1}^{I}\sum_{j=1}^{J}\left|\omega_{1(i,j)}-\omega_{2(i,j)}\right|^2}$
%     \end{adjustbox}
%     \label{eq:example}
% \end{equation}
\begin{equation}
    \resizebox{0.85\linewidth}{!}{
        $\mathcal{L}_{\mathrm{sharing}}\left(\omega_1,\omega_2\right)=\sqrt{\sum_{i=1}^{I}\sum_{j=1}^{J}\left|\omega_{1(i,j)}-\omega_{2(i,j)}\right|^2}$
    }
\end{equation}
where $\omega_1$ and $\omega_2$ represent the parameters of the two distinct models. Both models simultaneously share encoder parameters. Therefore, the loss function for the learning objective in soft parameter sharing is defined as:
% \begin{equation}
%     \begin{adjustbox}{max width=0.85\linewidth}
%         $Loss=(1-\lambda)\mathcal{L}_{\mathrm{cls\_loss\ }}+\lambda \mathcal{L}_{\mathrm{gen\_loss}}+\mathcal{L}_{\mathrm{sharing}}(\omega_1,\omega_2)$
%     \end{adjustbox}
%     \label{eq:example}
% \end{equation}
\begin{equation}
    \resizebox{0.85\linewidth}{!}{
        $Loss=(1-\lambda)\mathcal{L}_{\mathrm{cls\_loss}}+\lambda \mathcal{L}_{\mathrm{gen\_loss}}+\mathcal{L}_{\mathrm{sharing}}(\omega_1,\omega_2)$
    }
\end{equation}

% \paragraph{Intermediate Fine-Tuning~(INF)}
\noindentparagraph{\textbf{\textup{Intermediate Fine-Tuning~(INF).}}}
Using the intermediate fine-tuning approach, based on transfer learning principles~\cite {phang2018sentence}, our code ranking model is first trained on the generation task to understand the various reasons behind code discrepancies. It then undergoes further training focused on the classification objective.

For practical implementation, we opt for CodeT5+ (770M) as RankEF's base model, leveraging its state-of-the-art code pre-training capabilities and streamlined encoder-decoder structure that aligns with our minimal modification approach. In order for CodeT5+ to learn the ability to distinguish correct and incorrect code, we incorporate an MLP classification layer—comprising two linear layers and a ReLU activation—into CodeT5+'s encoder. Throughout our experimental analysis (See Section~\ref{section: RQ5}), we consistently set the multi-task training parameter, $\lambda$, at 0.9. Unless otherwise stated, subsequent experimental results refer to RankEF trained using the hard parameter sharing method, given its optimal performance. The training process of RankEF is illustrated in Algorithm ~\ref{alg:il}.

\begin{algorithm}[th!]
   \caption{RankEF's overall training process.}
   \label{alg:il}
\begin{algorithmic}[1]

   \State {\bfseries Input:} Code Gen. Dataset $(N, S)\in\mathcal{D}_{gen}$, initial Gen. LM $\pi_{\theta}$, initial Rank Model $\textsc{P}_{\theta}$, Complier $\textsc{EXEC}$, Execution Feedback Template $\mathcal{T}_{ef}$
   \State $\pi_{\theta^{*}} \leftarrow \textsc{FINETUNE}(\pi_{\theta}, \mathcal{D}_{gen})$ 
    \State \textcolor{blue}{// Sampling code using fine-tuned $\pi_{\theta^{*}}$}
   \State $C^* \leftarrow \pi_{\theta^{*}}(N\in D_{gen})$ 
   \State $\textsc{exection outcomes E} \leftarrow \textsc{EXEC}(S^*)$
   \State $\textsc{Execution feedback F} \leftarrow \mathcal{T}_{ef}(\textsc{EXEC}(C^*))$ 
   \State \textcolor{blue}{// Training data for RankEF}
   \State $D_{rank} \leftarrow \{N, S^*, E, F\}$ 
   \State \textcolor{blue}{// Hard Parameter Sharing}
   \For{$(N, S^*, E)\in D_{rank}\,\text{and}\,(N, S^*, F)\in D_{rank}$} 
   \State $\mathcal{L}_{cls\_loss} \leftarrow -\sum_i^n E_i \log P_{\theta}\left(\hat{E}_i \mid N, S^{*}\right)$
   \State $\mathcal{L}_{gen\_loss} \leftarrow -\sum \log P_{\theta}\left( F \mid N, S^{*}\right)$
   \State $\mathcal{L}_{total\_loss} \leftarrow (1-\lambda) \mathcal{L}_{cls\_loss} + \lambda \mathcal{L}_{gen\_loss}$
   \State $\textsc{Update}\,\,\,P_{\theta}\,\,\,with\,\,\,\mathcal{L}_{total\_loss}$
   \EndFor{}
   \State \textcolor{blue}{// Soft Parameter Sharing}
   \For{$(N, S^*, E)\in D_{rank}\,\text{and}\,(N, S^*, F)\in D_{rank}$} 
   \State $\mathcal{L}_{cls\_loss} \leftarrow -\sum_i^n E_i \log P_{\theta_{1}}\left(\hat{E}_i \mid N, C^{*}\right)$
   \State $\mathcal{L}_{gen\_loss} \leftarrow -\sum \log P_{\theta_{2}}\left( F \mid N, S^{*}\right)$
   \State $\mathcal{L}_{total\_loss} \leftarrow (1-\lambda) \mathcal{L}_{cls\_loss} + \lambda \mathcal{L}_{gen\_loss}+ \mathcal{L}_{\mathrm{sharing}}(\theta_1,\theta_2)$
   \State $\textsc{Update}\,\,\,P_{\theta_{1}}\,and\,P_{\theta_{2}}\,\,\,with\,\,\,\mathcal{L}_{total\_loss}$
   \EndFor{}
   \State \textcolor{blue}{// Intermediate Fine-Tuning(INF)}
   \For{$(N, S^*, F)\in D_{rank}$}
   \State $\mathcal{L}_{gen\_loss} \leftarrow -\sum \log P_{\theta}\left( F \mid N, S^{*}\right)$
   \State $P_{\theta^*} \leftarrow \textsc{Update}\,\,\,P_{\theta}\,\,\,with\,\,\,\mathcal{L}_{gen\_loss}$
   \EndFor{}
   \For{$(N, S^*, E)\in D_{rank}$}
   \State $\mathcal{L}_{cls\_loss} \leftarrow -\sum_i^n E_i \log P_{\theta}\left(\hat{E}_i \mid N, S^{*}\right)$
   \State $\textsc{Update}\,\,\,P_{\theta^*}\,\,\,with\,\,\,\mathcal{L}_{cls\_loss}$
   \EndFor{}
\end{algorithmic}
\end{algorithm}

\subsection{Final Ranking Process}

Here, we outline RankEF's final inference process, depicted in phase \textbf{C} of Figure~\ref{fig: model}. This procedure encompasses two distinct stages: code sampling and subsequent ranking. As delineated in earlier sections, RankEF, using its model $P$ and the formatted input $CLS\_S$, computes a probability score $s = P(Correct|CLS\_S)$. This score emerges from the softmax transformation of the hidden states yielded by an appended MLP classification layer. Notably, only execution feedback is utilized as training data in this phase, eliminating the need for such feedback during final ranking.

\section{Experimental Setup}
\subsection{Research Questions}
\textbf{RQ1: Overall Performance.} Can RankEF perform well compared to single-task approach?

\noindent\textbf{Motivation.} As previously mentioned, the current work on neural network-based code ranking considers code ranking as a single-task code classification. In response, we introduce execution feedback as an additional supervisory signal. Therefore, we have formulated this RQ to study how RankEF performs on complex programming problems and programming problems of varying difficulties, compared to single-task approach.

\noindent\textbf{RQ2: Transferability.} How is RankEF's transferability?

\noindent\textbf{Motivation.} Since the training data used by RankEF comes from APPS, an evaluation of RankEF's transfer ability on other datasets is necessary. Therefore, we formulate this RQ to investigate the migration ability of RankEF on MBPP and HumanEval benchmarks.

\noindent\textbf{RQ3: Compare to other non-executive methods.} How does the RankEF perform compared to other non-execution methods?

\noindent\textbf{Motivation.} There are many non-execution-based code ranking methods other than CodeRanker. We feel it is necessary to compare RankEF with these methods to explore whether RankEF can have stronger performance.

\noindent\textbf{RQ4: Multi-Task Training Effectiveness.} How do different multi-task training methods affect RankEF differently?

\noindent\textbf{Motivation.} 
% Since RankEF is trained using multi-task training methods, different multi-task training methods may have different effects on RankEF, so it is necessary to conduct an in-depth study on RankEF for different multi-task methods. In this regard, we investigate the effects of three different multi-task methods on RankEF.
Given RankEF's use of multi-task training, it's essential to explore its response to various methods. We investigate the effects of three different multi-task approaches on RankEF.

\noindent\textbf{RQ5: Task Weighting Parameters.} What is the impact of the task weighting parameters for RankEF?

\noindent\textbf{Motivation.} As the RankEF method involves two key tasks: code classification and execution feedback prediction, researches~\cite{sener2018multi, vandenhende2021multi} in the field of multitask learning have shown that there may be conflicting phenomena between different tasks, which can lead to a decrease in the effectiveness of multitask learning. Adjusting parameter weights is crucial to balance these tasks' impacts on learning outcomes, necessitating an investigation into RankEF's performance under varying weights.

\subsection{Datasets and Metrics}
We chose three widely-used code generation datasets in our experiments.

\setlist[itemize]{left=0pt}
\begin{itemize}
\item \textbf{APPS} \cite{hendrycks2021measuring} compiles 10,000 problems from platforms like Codeforces and LeetCode, divided into 5,000 each for training and testing. The problems are categorized into introductory, interview, and competition levels. Following previous work \cite{inala2022fault, zhang-etal-2023-self}, the same set of 598 problems is used for validation during training, while the rest is used for training.

\item \textbf{MBPP} \cite{austin2021program} includes 974 crowdsourced programming problems, allocated as 374 for training, 90 for validation, 500 for testing, and 10 for few-shot prompt learning. In this paper, we only use 500 test set problems to evaluate the transferability of RankEF.

\item \textbf{HumanEval}~\cite{humaneval} consists of 164 artificially constructed basic programming questions, including function signatures, docstrings, and multiple unit tests.
\end{itemize}
Drawing upon prior studies, we utilize the Pass@$k$ metric \cite{humaneval} for model accuracy evaluation. In our experimental setup, we sample a set of 100 code snippets for each problem. To assess the effectiveness of RankEF and CodeRanker, we calculate Pass@\{1, 2, 5\} metrics for the top-ranked 1, 2, and 5 code by these tools. In scenarios where no ranking mechanism is used, we randomly select 1, 2, and 5 code out of the 100-sample pool to compute the Pass@\{1, 2, 5\} metrics.

\begin{table*}[htbp]
\caption{Results on APPS validation dataset.}
\vspace{-1em}
\resizebox{1.0\textwidth}{!} {
\begin{tabular}{cc|ccc|ccc|ccc}
\hline
                        &                        & \multicolumn{3}{c|}{Pass@1}                                                               & \multicolumn{3}{c|}{Pass@2}                                                               & \multicolumn{3}{c}{Pass@5}                                                                \\ \cline{3-11} 
\multirow{-2}{*}{Model} & \multirow{-2}{*}{para} & Random & CodeRanker* & RankEF         & Random & CodeRanker* & RankEF         & Random & CodeRanker* & RankEF         \\ \hline
PyCodeGPT               & 110M                   & 4.52       & 6.99        & \cellcolor[HTML]{DDDDDD}\textbf{9.53}  & 7.02       & 10.85        & \cellcolor[HTML]{DDDDDD}\textbf{13.04} &    12.04      & 16.08       & \cellcolor[HTML]{DDDDDD}\textbf{19.23} \\
StarCoder               & 164M                   & 6.41       & 9.86        & \cellcolor[HTML]{DDDDDD}\textbf{12.04} & 10.11      & 13.23       & \cellcolor[HTML]{DDDDDD}\textbf{15.34} & 17.22      & 19.76       & \cellcolor[HTML]{DDDDDD}\textbf{21.88} \\
CodeGen                 & 350M                   & 6.69       & 11.28       & \cellcolor[HTML]{DDDDDD}\textbf{15.91} & 11.87      & 15.51       & \cellcolor[HTML]{DDDDDD}\textbf{19.77} & 19.57      & 24.66       & \cellcolor[HTML]{DDDDDD}\textbf{28.64} \\
CodeT5+                 & 770M                   & 12.54      & 16.42       & \cellcolor[HTML]{DDDDDD}\textbf{18.46} & 18.42      & 22.13       & \cellcolor[HTML]{DDDDDD}\textbf{23.87} & 27.76      & 31.08       & \cellcolor[HTML]{DDDDDD}\textbf{33.93} \\
GPT-Neo                 & 1.3B                   & 6.06       & 9.39        & \cellcolor[HTML]{DDDDDD}\textbf{11.54}  & 9.88       & 12.87       & \cellcolor[HTML]{DDDDDD}\textbf{15.22} & 15.12      & 19.04       & \cellcolor[HTML]{DDDDDD}\textbf{22.07} \\
CodeGen                 & 2B                     & 15.72      &  19.23      & \cellcolor[HTML]{DDDDDD}\textbf{25.08} & 24.41      & 27.92       & \cellcolor[HTML]{DDDDDD}\textbf{31.61} & 34.39      & 37.29       & \cellcolor[HTML]{DDDDDD}\textbf{41.14}  \\
CodeLlama                 & 7B                     & 18.56      &  25.10      & \cellcolor[HTML]{DDDDDD}\textbf{28.93} & 29.26      & 34.49       & \cellcolor[HTML]{DDDDDD}\textbf{38.13} & 43.48      & 48.84       & \cellcolor[HTML]{DDDDDD}\textbf{53.68}  \\ \hline
\end{tabular}
}
\label{tab:tabel_1}
\vspace{-1em}
\end{table*}
\begin{table*}[t]
\caption{Results on APPS Test Dataset (Solved Problems). StarCoder (15B) and GPT-3.5, are not fine-tuned on the APPS dataset. Instead, they generate code candidates in a one-shot manner (with GPT-3.5 using only the most difficult 1,000 problems in the test set) and utilize rankers trained on samples from other generative models (second row). `StarCoder-Best' and `GPT-3.5-Turbo-Best' signify the peak performance achieved by these diverse rankers.
} 
\vspace{-1em}
\resizebox{1.0\textwidth}{!} {
\begin{tabular}{cc|ccc|ccc|ccc}
\hline
                        &                        & \multicolumn{3}{c|}{Pass@1}                                                               & \multicolumn{3}{c|}{Pass@2}                                                               & \multicolumn{3}{c}{Pass@5}                                                                \\ \cline{3-11} 
\multirow{-2}{*}{Model} & \multirow{-2}{*}{para} & Random & CodeRanker* & RankEF         & Random & CodeRanker* & RankEF         & Random & CodeRanker* & RankEF         \\ \hline
PyCodeGPT               & 110M                   & 6.07       & 12.14         & \cellcolor[HTML]{DDDDDD}\textbf{18.85}  & 8.94        & 17.57       & \cellcolor[HTML]{DDDDDD}\textbf{25.56} & 19.81      & 31.31       & \cellcolor[HTML]{DDDDDD}\textbf{40.26}  \\
StarCoder               & 164M                   & 6.22       & 11.95        & \cellcolor[HTML]{DDDDDD}\textbf{13.21}  & 13.27       & 16.62       & \cellcolor[HTML]{DDDDDD}\textbf{19.88} & 26.23      & 30.57       & \cellcolor[HTML]{DDDDDD}\textbf{34.20} \\
CodeGen                 & 350M                   & 8.07       & 12.08        & \cellcolor[HTML]{DDDDDD}\textbf{15.47}  & 13.67      & 18.39        & \cellcolor[HTML]{DDDDDD}\textbf{25.34} & 24.44      & 32.73       & \cellcolor[HTML]{DDDDDD}\textbf{38.16}  \\
CodeT5+                 & 770M                   & 12.42      & 15.82       & \cellcolor[HTML]{DDDDDD}\textbf{19.76} & 21.05      & 25.32       & \cellcolor[HTML]{DDDDDD}\textbf{31.29} & 35.06      & 39.52       & \cellcolor[HTML]{DDDDDD}\textbf{45.18} \\
GPT-Neo                 & 1.3B                   & 6.82       & 9.49  & \cellcolor[HTML]{DDDDDD}\textbf{11.45}      & 11.67      & 14.68       & \cellcolor[HTML]{DDDDDD}\textbf{17.21}      & 22.67      & 25.63       & \cellcolor[HTML]{DDDDDD}\textbf{29.88}      \\
CodeGen                 & 2B                     & 10.48      & 13.96       & \cellcolor[HTML]{DDDDDD}\textbf{17.50} & 17.66      & 20.69       & \cellcolor[HTML]{DDDDDD}\textbf{26.62} & 32.84      & 33.93       & \cellcolor[HTML]{DDDDDD}\textbf{41.93}  \\  
CodeLlama                 & 7B                     & 15.03      & 18.02       & \cellcolor[HTML]{DDDDDD}\textbf{23.12} & 24.47      & 26.59       & \cellcolor[HTML]{DDDDDD}\textbf{33.04} & 38.25      & 43.26       & \cellcolor[HTML]{DDDDDD}\textbf{47.98}  \\ \hline
StarCoder-Best                 & 15B                     & 4.90      & 9.95       & \cellcolor[HTML]{DDDDDD}\textbf{16.60} & 8.11      & 20.03       & \cellcolor[HTML]{DDDDDD}\textbf{29.60} & 17.35      & 38.26       & \cellcolor[HTML]{DDDDDD}\textbf{50.19}  \\ 
GPT-3.5-Turbo-Best                 & -                     & 4.26      & 8.51       & \cellcolor[HTML]{DDDDDD}\textbf{10.11} & 9.04      & 13.83       & \cellcolor[HTML]{DDDDDD}\textbf{19.15} & 28.72      & 31.41       & \cellcolor[HTML]{DDDDDD}\textbf{38.30}  \\ \hline
\end{tabular}
}
\label{tab:tabel_2}
\vspace{-1em}
\end{table*}

\subsection{Base Generation Models and Baselines} 
In this paper, we employ a diverse array of leading code pre-training models, such as PyCodeGPT (110M) \cite{CERT}, StarCoder (164M, 15B) \cite{li2023starcoder}, CodeT5+ (770M) \cite{wang2023codet5+}, GPT-Neo (1.3B) \cite{black2021gpt}, CodeGen (350M, 2B) \cite{nijkamp2022codegen}, CodeLlama (7B) \cite{codellama}, and GPT-3.5-Turbo~\cite{ChatGPT}. 
% Ranging from 110M to 175B in parameters, these models are chosen to showcase the efficacy and scalability of our proposed approach. 
% All models, except StarCoder (15B) and GPT-3.5-Turbo, undergo fine-tuning on the APPS dataset. 
These models are evaluated under fine-tune or non-fine-tune experimental settings, detailed explanations can be found in Tables ~\ref{tab:tabel_2}, ~\ref{tab:tabel_4}, and ~\ref{tab:tabel_5}.
% \hy{finetun or fine-tune?} \zh{done.}}

We assess the performance of RankEF against the neural-network-based CodeRanker \cite{inala2022fault} when applied to these base generation models. Our evaluation of CodeRanker's efficacy is grounded on tests conducted on our curated dataset, training it with the Ternary task as suggested in the original study. This choice stems from the insight that CodeRanker exhibits enhanced performance with models adept in robust code generation when trained on the Ternary task. We refer to this tailored version of CodeRanker as CodeRanker*. Notably, in all experiments, RankEF and CodeRanker both use the same underlying ranker model, CodeT5+. In addition to CodeRanker, we chose other non-execution-based code ranking methods, including Coder~\cite{humaneval}, Coder-Reviewer~\cite{zhang2023coder}, and their variants.

% \subsection{Evaluation Metrics.} 
% Drawing upon prior studies, we utilize the Pass@$k$ metric \cite{humaneval} for model accuracy evaluation. For each problem sampled to generate n>=k copies of code, the number of correct codes c<=n, pass@k metric is calculated as follows:
% \begin{equation}
% \operatorname{pass} @ k=\underset{\text { Problems }}{\mathbb{E}}\left[1-\frac{\left(\begin{array}{c}
% n-c \\
% k
% \end{array}\right)}{\left(\begin{array}{l}
% n \\
% k
% \end{array}\right)}\right]
% \end{equation}
% In our experimental setup, we sample a set of 100 code snippets for each problem. To assess the effectiveness of RankEF and CodeRanker, we calculate Pass@\{1, 2, 5\} metrics for the top-ranked 1, 2, and 5 codes by these tools. In scenarios where no ranking mechanism is used, we randomly select 1, 2, and 5 codes out of the 100-sample pool to compute the Pass@\{1, 2, 5\} metrics.

\subsection{Training/Inference Settings} 
To ensure data diversity in training RankEF, we adopt the strategies from previous researches ~\cite{inala2022fault, cobbe2021training}, fine-tuning the base generation models over 2 epochs at a learning rate of 2e-5. To get a diversity of code candidates, we sample code with a temperature of 0.8 and a top-$p$ setting of 0.95. For RankEF, given its larger ranking dataset compared to the code generation set, we train up to 5 epochs at a 1e-4 learning rate and a 256 batch size, choosing the optimal checkpoint based on validation results. All experiments are conducted using 6 RTX A6000-48GB GPUs.
\section{Experimental Results}

We conduct several experiments to answer our research questions. In Section~\ref{section: RQ1}, we focus on evaluating RankEF's overall performance. In Section~\ref{section: RQ2}, we evaluate RankEF's transfer ability on the MBPP and HumanEval benchmarks. In Section~\ref{section: RQ3}, we compare RankEF with other non-execution code ranking methods. In Sections~\ref{section: RQ4} and~\ref{section: RQ5}, we analyze the impact of different multi-task methods and the impact of different task weights, respectively.

\begin{table}[t]
\caption{Results on all APPS test sets (Solved Problems) on three topics of varying difficulty. CodeGen (350M), after fine-tuning, serves as the base generation model. The red percentage indicates the relative improvement of the metric after ranking compared to that before ranking.}
\vspace{-1em}
\resizebox{0.48\textwidth}{!} {
\begin{tabular}{cccccccc}
\hline
Difficulty level        & Method                              & \multicolumn{2}{c}{Pass@1}                                                           & \multicolumn{2}{c}{Pass@2}                                                           & \multicolumn{2}{c}{Pass@5}                                                           \\ \hline
                        & Random                          & 8.47                         &                                                       & 16.40                         &                                                       & 26.46                         &                                                       \\
                        & CodeRanker* & 16.40 & {\color[HTML]{FE0000} 93.6\%} & 23.81 & {\color[HTML]{FE0000} 45.2\%} & 38.10 & {\color[HTML]{FE0000} 44.0\%} \\
\multirow{-3}{*}{Intro} & \cellcolor[HTML]{DDDDDD}RankEF      & \cellcolor[HTML]{DDDDDD}\textbf{18.52} & \cellcolor[HTML]{DDDDDD}{\color[HTML]{FE0000} 118.7\%} & \cellcolor[HTML]{DDDDDD}\textbf{26.99} & \cellcolor[HTML]{DDDDDD}{\color[HTML]{FE0000} 64.6\%} & \cellcolor[HTML]{DDDDDD}\textbf{41.80} & \cellcolor[HTML]{DDDDDD}{\color[HTML]{FE0000} 58.0\%} \\ \hline
                        & Random                          & 6.97                         &                                                       & 12.98                         &                                                       & 24.52                         &                                                       \\
                        & CodeRanker* & 10.34 & {\color[HTML]{FE0000} 48.4\%} & 15.87 & {\color[HTML]{FE0000} 22.3\%} & 31.73 & {\color[HTML]{FE0000} 29.4\%} \\
\multirow{-3}{*}{Inter} & \cellcolor[HTML]{DDDDDD}RankEF      & \cellcolor[HTML]{DDDDDD}\textbf{14.90} & \cellcolor[HTML]{DDDDDD}{\color[HTML]{FE0000} 113.8\%} & \cellcolor[HTML]{DDDDDD}\textbf{21.63} & \cellcolor[HTML]{DDDDDD}{\color[HTML]{FE0000} 66.6\%} & \cellcolor[HTML]{DDDDDD}\textbf{36.06} & \cellcolor[HTML]{DDDDDD}{\color[HTML]{FE0000} 47.1\%} \\ \hline
                        & Random                          & 2.72                         &                                                       & 5.44                         &                                                       & 13.61                         &                                                       \\
                        & CodeRanker* & 3.06 & {\color[HTML]{FE0000} 12.5\%} & 6.12 & {\color[HTML]{FE0000} 12.5\%} & 15.31 & {\color[HTML]{FE0000} 12.5\%} \\
\multirow{-3}{*}{Comp}  & \cellcolor[HTML]{DDDDDD}RankEF      & \cellcolor[HTML]{DDDDDD}\textbf{5.10} & \cellcolor[HTML]{DDDDDD}{\color[HTML]{FE0000} 87.5\%}  & \cellcolor[HTML]{DDDDDD}\textbf{9.80} & \cellcolor[HTML]{DDDDDD}{\color[HTML]{FE0000} 80.1\%}  & \cellcolor[HTML]{DDDDDD}\textbf{21.43} & \cellcolor[HTML]{DDDDDD}{\color[HTML]{FE0000} 57.5\%} \\ \hline
\end{tabular}
}
%We use the fine-tuned CodeGen-350M as the base generation model.}
\label{tab:tabel_3}
\vspace{-1em}
\end{table}
\subsection{RQ1: Overall Performance}
\label{section: RQ1}
\textbf{APPS Validation Dataset.} Our initial evaluation of RankEF is conducted on the validation set of the APPS dataset. Table~\ref{tab:tabel_1} shows the results of RankEF, trained through multi-task paradigm, in comparison to CodeRanker, trained via singular task paradigm, across seven distinct models on the aforementioned APPS validation set. In juxtaposition with the singular task-trained CodeRanker, RankEF manifests a considerable augmentation in the Pass@\{1,2,5\} metrics. This suggests that a multi-task training methodology holds a pronounced advantage over its single-task counterpart.
\begin{table*}[t]
\caption{Results of the transfer capabilities on MBPP. `Finetuned on APPS' refers to using base generation models fine-tuned on APPS to sample candidate code, while `Non-fine-tuned' refers to using base models that have not been fine-tuned on APPS.}
\vspace{-1em}
\resizebox{1.0\textwidth}{!} {
\begin{tabular}{cc|ccc|ccc|ccc}
\hline
                        &                        & \multicolumn{3}{c|}{Pass@1}                                                               & \multicolumn{3}{c|}{Pass@2}                                                               & \multicolumn{3}{c}{Pass@5}                                                                \\ \cline{3-11} 
\multirow{-2}{*}{Model} & \multirow{-2}{*}{para} & Random & CodeRanker* & RankEF         & Random & CodeRanker* & RankEF         & Random & CodeRanker* & RankEF         \\ \hline
\multicolumn{2}{c|}{\textit{\textbf{Fine-tuned on APPS}}}          &        &             &        &        &             &        &        &             &        \\
PyCodeGPT               & 110M                   & 5.84       & 8.10        & \cellcolor[HTML]{DDDDDD}\textbf{13.80}  & 9.92       & 12.96       & \cellcolor[HTML]{DDDDDD}\textbf{17.98} & 16.20      & 19.23       & \cellcolor[HTML]{DDDDDD}\textbf{24.00}  \\
StarCoder               & 164M                   & 6.77       & 8.52        & \cellcolor[HTML]{DDDDDD}\textbf{13.80} & 11.98      & 14.47       & \cellcolor[HTML]{DDDDDD}\textbf{18.00} & 18.22      & 20.02       & \cellcolor[HTML]{DDDDDD}\textbf{24.29} \\
CodeGen                 & 350M                   & 13.94      & 16.24       & \cellcolor[HTML]{DDDDDD}\textbf{19.12} & 20.12      & 22.48       & \cellcolor[HTML]{DDDDDD}\textbf{24.88} & 29.00      & 31.24       & \cellcolor[HTML]{DDDDDD}\textbf{33.63} \\
CodeT5+                 & 770M                   & 18.28      & 21.44       & \cellcolor[HTML]{DDDDDD}\textbf{24.28} & 25.94      & 27.01       & \cellcolor[HTML]{DDDDDD}\textbf{30.02} & 35.60      & 38.42       & \cellcolor[HTML]{DDDDDD}\textbf{40.01} \\
GPT-Neo                 & 1.3B                   & 6.42       & 10.04       & \cellcolor[HTML]{DDDDDD}\textbf{11.68} & 11.84      & 13.70       & \cellcolor[HTML]{DDDDDD}\textbf{16.20}  & 18.41     & 19.20       & \cellcolor[HTML]{DDDDDD}\textbf{21.02} \\
CodeGen                 & 2B                     & 26.45      & 29.12       & \cellcolor[HTML]{DDDDDD}\textbf{31.85} & 36.36      & 38.03            & \cellcolor[HTML]{DDDDDD}\textbf{39.56} & 46.80      & 47.20       & \cellcolor[HTML]{DDDDDD}\textbf{49.47}  \\ 
CodeLlama               & 7B                     & 38.40      & 40.02       & \cellcolor[HTML]{DDDDDD}\textbf{41.98} & 46.30      & 48.38            & \cellcolor[HTML]{DDDDDD}\textbf{49.74} & 58.80      & 58.91       & \cellcolor[HTML]{DDDDDD}\textbf{59.72}  \\ \hline
\multicolumn{2}{c|}{\textit{\textbf{Non-fine-tuned}}}          &        &             &        &        &             &        &        &             &        \\
CodeT5+                 & 770M                   & 12.00       & 14.72       & \cellcolor[HTML]{DDDDDD}\textbf{16.60} & 18.60      & 21.10       & \cellcolor[HTML]{DDDDDD}\textbf{22.92}  & 28.00     & 31.00       & \cellcolor[HTML]{DDDDDD}\textbf{32.20} \\
CodeGen                 & 2B                     & 24.96      & 28.95       & \cellcolor[HTML]{DDDDDD}\textbf{30.78} & 33.19      & 35.01            & \cellcolor[HTML]{DDDDDD}\textbf{36.27} & 43.58      & 44.06       & \cellcolor[HTML]{DDDDDD}\textbf{45.66}  \\ 
CodeLlama               & 7B                     & 34.00      & 35.65       & \cellcolor[HTML]{DDDDDD}\textbf{38.36} & 44.86      & 45.33            & \cellcolor[HTML]{DDDDDD}\textbf{46.92} & 56.33      & 58.01       & \cellcolor[HTML]{DDDDDD}\textbf{59.11}  \\ \hline
\end{tabular}
}
\label{tab:tabel_4}
\vspace{-1em}
\end{table*}
\begin{table*}[t]
\caption{Results of the transfer capabilities on HumanEval.}
\vspace{-1em}
\resizebox{1.0\textwidth}{!} {
\begin{tabular}{cc|ccc|ccc|ccc}
\hline
                        &                        & \multicolumn{3}{c|}{Pass@1}                                                               & \multicolumn{3}{c|}{Pass@2}                                                               & \multicolumn{3}{c}{Pass@5}                                                                \\ \cline{3-11} 
\multirow{-2}{*}{Model} & \multirow{-2}{*}{para} & Random & CodeRanker* & RankEF         & Random & CodeRanker* & RankEF         & Random & CodeRanker* & RankEF         \\ \hline
\multicolumn{2}{c|}{\textit{\textbf{Fine-tuned on APPS}}}           &        &             &        &        &             &        &        &             &        \\
PyCodeGPT               & 110M                   & 4.76      & 6.10        & \cellcolor[HTML]{DDDDDD}\textbf{8.17}  & 7.26       & 8.72        & \cellcolor[HTML]{DDDDDD}\textbf{10.67}  & 10.98      & 12.20       & \cellcolor[HTML]{DDDDDD}\textbf{13.41}  \\
StarCoder               & 164M                   & 6.34       & 7.20        & \cellcolor[HTML]{DDDDDD}\textbf{8.79}  & 7.99       & 8.78        & \cellcolor[HTML]{DDDDDD}\textbf{10.26} & 10.37      & 11.02       & \cellcolor[HTML]{DDDDDD}\textbf{12.20} \\
CodeGen                 & 350M                   & 6.71       & 7.20       & \cellcolor[HTML]{DDDDDD}\textbf{9.39}  & 8.78       & 10.30       & \cellcolor[HTML]{DDDDDD}\textbf{11.46} & 10.98      & 13.41       & \cellcolor[HTML]{DDDDDD}\textbf{15.24} \\
CodeT5+                 & 770M                   & 10.89      & 12.89       & \cellcolor[HTML]{DDDDDD}\textbf{13.54} & 14.24      & 15.84       & \cellcolor[HTML]{DDDDDD}\textbf{16.59} & 19.51      & 21.34       & \cellcolor[HTML]{DDDDDD}\textbf{23.07} \\
GPT-Neo                 & 1.3B                   & 4.76       & 6.08   & \cellcolor[HTML]{DDDDDD}\textbf{7.21}  & 6.34       & 7.88        & \cellcolor[HTML]{DDDDDD}\textbf{8.78}  & 7.92       & 10.98       & \cellcolor[HTML]{DDDDDD}\textbf{12.20}      \\
CodeGen                 & 2B                     & 14.63      & 17.32       & \cellcolor[HTML]{DDDDDD}\textbf{19.39} & 20.06      & 22.99       & \cellcolor[HTML]{DDDDDD}\textbf{24.94} & 27.44      & 32.93       & \cellcolor[HTML]{DDDDDD}\textbf{34.76}  \\ 
CodeLlama                 & 7B                     & 23.29      & 25.90       & \cellcolor[HTML]{DDDDDD}\textbf{29.39} & 32.20      & 34.07       & \cellcolor[HTML]{DDDDDD}\textbf{37.99} & 45.12      & 44.46       & \cellcolor[HTML]{DDDDDD}\textbf{48.17}  \\ \hline
\multicolumn{2}{c|}{\textit{\textbf{Non-fine-tuned}}}          &        &             &        &        &             &        &        &             &        \\
CodeT5+                 & 770M                   & 11.34       & 12.56   & \cellcolor[HTML]{DDDDDD}\textbf{14.63}  & 15.61       & 17.11        & \cellcolor[HTML]{DDDDDD}\textbf{19.26}  & 21.34       & 23.43       & \cellcolor[HTML]{DDDDDD}\textbf{25.00}      \\
CodeGen                 & 2B                     & 17.80      & 19.63       & \cellcolor[HTML]{DDDDDD}\textbf{22.86} & 24.18      & 25.11       & \cellcolor[HTML]{DDDDDD}\textbf{28.78} & 32.53      & 34.21       & \cellcolor[HTML]{DDDDDD}\textbf{36.59}  \\ 
CodeLlama                 & 7B                     & 25.00      & 27.37       & \cellcolor[HTML]{DDDDDD}\textbf{30.49} & 34.69      & 36.27       & \cellcolor[HTML]{DDDDDD}\textbf{38.92} & 48.17      & 47.63       & \cellcolor[HTML]{DDDDDD}\textbf{51.27}  \\ \hline
\end{tabular}
}

\label{tab:tabel_5}
\vspace{-1em}
\end{table*}

\noindent\textbf{APPS Test Dataset~(Solved Problems).} We further subject RankEF to an evaluation on the APPS test set, mirroring the assessment criteria employed on the validation set. It is noteworthy that the test set of APPS presents a significantly elevated level of complexity compared to the validation set, coupled with a more extensive set of unit tests. This intricate nature results in rather diminutive metric values when assessing code generation models on the test set. Recognizing that code ranking does not fundamentally augment the intrinsic code generation capability of a model (a set of entirely erroneous code will remain erroneous post-ranking), our assessment of the test set focuses on problems that the generation models could feasibly address. This metric is determined by the criterion that, out of 100 code samples, at least one meets the stipulated requirements, thus more accurately reflecting the efficacy of code ranking. 

Table~\ref{tab:tabel_2} offers insights into the performance of both CodeRanker and RankEF across the nine distinct code generation models on solvable problems within the APPS test set. Observationally, RankEF, when trained via a multi-task paradigm, consistently outperforms the singular task-trained CodeRanker across the Pass@\{1,2,5\} metrics for various models. Moreover, it is discerned that RankEF enables models with smaller parameter volumes to achieve performance metrics on par with models possessing several times their parameter magnitude. Given that larger models entail higher fine-tuning costs, RankEF, to a certain extent, indirectly mitigates the expenses associated with fine-tuning. 

\noindent\textbf{APPS Different Difficulty-Level Problems.}  Table~\ref{tab:tabel_3} shows the results of RankEF with multi-task training and CodeRanker with single-task training on CodeGen (2B) for the APPS test set at three different difficulty levels~(Introductory, Interview, Competition). We observe that RankEF shows stronger performance on more difficult problems; as the difficulty level increases, RankEF's advantage over CodeRanker becomes more and more obvious. This suggests that the introduction of execution feedback allows RankEF to truly understand why the code is going wrong, and thus RankEF has better performance on harder problems.

% \begin{tcolorbox}[size=title]
% \textbf{Answer to RQ1:} \textcolor{black}{Compared to the single-task CodeRanker, RankEF demonstrates superior performance across various base models, irrespective of the parameter size (110M$\sim$175B). Moreover, under different difficulty settings, RankEF has more obvious advantages on the most difficult problems.}
% \end{tcolorbox}

\find{\textbf{\ding{45} Answer to RQ1:} \textcolor{black}{Compared to the single-task CodeRanker, RankEF demonstrates superior performance across various base models, irrespective of the parameter size. Moreover, under different difficulty settings, RankEF has more obvious advantages on the most difficult problems.}}

\subsection{RQ2: Transferability}
\label{section: RQ2}
In addition to APPS, we evaluate RankEF's transfer capabilities on two other prominent code generation benchmarks, namely MBPP and HumanEval. To ensure a fair comparison, we follow the methodology of CodeRanker~\cite{inala2022fault} by using base generation models fine-tuned on the APPS dataset in our experiments. However, this approach carries the risk of overfitting to the APPS-specific coding style. To mitigate this, we also conducted experiments with several larger base generation models that have not been fine-tuned on APPS. For the MBPP dataset, since it provides data for few-shot prompt learning, we follow previous research and sample code in a few-shot manner~\cite{wang2023codet5+, codellama}. Specifically, we select one example as prompt to sample code in one-shot manner. In contrast, for the HumanEval dataset, which does not have additional data, we sample code using a zero-shot approach, meaning no example is used as prompts.

As indicated in Table~\ref{tab:tabel_4} and Table~\ref{tab:tabel_5}, it is observed that RankEF, compared to the singular task-trained CodeRanker, consistently manifests superior transfer capabilities on both MBPP and HumanEval benchmarks. Compared to the singular task-trained CodeRanker, RankEF demonstrates superior transfer capabilities on both MBPP and HumanEval benchmarks. Additionally, we observed that fine-tuning on the APPS dataset improves the performance of base generation models on MBPP but diminishes their performance on HumanEval. This phenomenon can be attributed to different task setups: HumanEval requires code completion based on partial code snippets, unlike MBPP and APPS, which involve generating complete code from natural language descriptions. Fine-tuned base generation models on APPS tend to generate entire code segments from scratch on HumanEval, rather than completing the given partial snippets, which increases the task's complexity. In contrast, models that are not fine-tuned on APPS can complete the code based on the provided snippets, reducing the task's complexity. Nevertheless, RankEF consistently demonstrates stronger transfer capabilities on both MBPP and HumanEval, regardless of whether the base generation models are fine-tuned or not.

% \begin{tcolorbox}[size=title]
% \textbf{Answer to RQ2:} \textcolor{black}{We trained only on APPS and performed migration experiments on MBPP and HumanEval. RankEF consistently shows superior transfer capabilities on both MBPP and HumanEval, regardless of whether the base generation models have been fine-tuned or not.}
% \end{tcolorbox}

\find{
\textbf{\ding{45} Answer to RQ2:} \textcolor{black}{We trained only on APPS and performed migration experiments on MBPP and HumanEval. RankEF consistently shows superior transfer capabilities on both MBPP and HumanEval, regardless of whether the base generation models have been fine-tuned or not.}
}

\subsection{RQ3: Compare to Non-Executive Methods}
\label{section: RQ3}

We also compare RankEF with other non-executive code ranking methods~\cite{humaneval,zhang2023coder}. We report the experimental results of RankEF with a range of non-executive methods on APPS and MBPP as shown in Table~\ref{tab: tabel_7}. For the APPS dataset, we use the CodeGen-2B model fine-tuned on APPS, while for the MBPP dataset, we use the same base CodeGen-2B model as mentioned in the original paper~\cite{zhang2023coder}.

For the APPS benchmark, Coder-Reviewer's original article does not conduct experiments on this benchmark, and we find that the methodology Coder-Reviewer arrived at does not cope with this programming competition type of complexity as it is very difficult to reason backwards about complex programming contest problem descriptions based on code. We followed its methodology and conducted relevant experiments on the APPS benchmark to verify our point. The experimental results show that RankEF is consistently and significantly better than other baselines on the APPS benchmark. For MBPP, we report results from the original paper~\cite{zhang2023coder} under the same model as well as our experimental results, which show that RankEF outperforms most methods with comparable or even better performance.

% \begin{tcolorbox}[size=title]
% \textbf{Answer to RQ3:} \textcolor{black}{Compared to other non-execution code ranking methods, RankEF has a solid advantage on the most challenging APPS benchmark, and also shows the best performance on the simpler MBPP benchmark.}
% \end{tcolorbox}

\find{
\textbf{\ding{45} Answer to RQ3:} \textcolor{black}{Compared to other non-execution code ranking methods, RankEF has a solid advantage on the most challenging APPS benchmark, and also shows the best performance on the simpler MBPP benchmark.}
}

\begin{table*}[t]
\caption{Results on comparing with other non-executive code ranking methods. Where `Random' stands for no ranking method, `Coder' is the method proposed by Chen et al.~\cite{humaneval}, and the rest of the methods are proposed by Zhang et al.~\cite{zhang2023coder}.}
\vspace{-1em}
\resizebox{0.96\textwidth}{!} {
\begin{tabular}{cccccccc
>{\columncolor[HTML]{DDDDDD}}c }
\hline
\multicolumn{2}{c}{Method}                                                         & Random & Reviewer & Coder & Coder-Reviewer & N.Coder & N.Coder-Reviewer & RankEF         \\ \hline
\multicolumn{2}{c}{APPS}                                                           & 10.48   & 9.10     & 10.34  & 11.03           & 7.44    & 7.59             & \textbf{17.50}  \\ \hline
                       & Our Results                                               & 24.96  & -        & -     & -              & -       & -                & \textbf{30.78} \\ \cline{2-2}
\multirow{-2}{*}{MBPP} & \begin{tabular}[c]{@{}c@{}}Reported in \\ ~\citep{zhang2023coder}\end{tabular} & 24.10   & 28.40     & 28.90  & 30.50           & 27.90    & 29.30             & -              \\ \hline
\end{tabular}
}

\label{tab: tabel_7}
\vspace{-1em}
\end{table*}
\begin{table}[t]
\caption{Results of different multi-task training methods on APPS, where $r$ represents the number of rounds of RankEF-INF. CodeGen (2B), after fine-tuning, serves as the base generation model.}
\vspace{-1em}
\resizebox{0.48\textwidth}{!} {
\begin{tabular}{cccccccc}
\hline
\multicolumn{2}{c}{}                                    & \multicolumn{3}{c}{Val}                          & \multicolumn{3}{c}{Test~(Solved Problems)}        \\ \cline{3-8} 
\multicolumn{2}{c}{\multirow{-2}{*}{Methods}}           & Pass@1         & Pass@2         & Pass@5         & Pass@1         & Pass@2         & Pass@5         \\ \hline
\multicolumn{2}{c}{Random}                              & 15.72          & 24.41          & 34.39          & 10.48          & 17.66          & 32.84          \\
\multicolumn{2}{c}{CodeRanker*}                         & 19.23          & 27.92          & 37.29          & 13.96          & 20.69          & 33.93          \\ \hline
\rowcolor[HTML]{DDDDDD} 
\multicolumn{2}{c}{\cellcolor[HTML]{DDDDDD}RankEF-Hard} & \textbf{25.08} & \textbf{31.61} & \textbf{41.14} & \textbf{17.50} & \textbf{26.62} & \textbf{41.93} \\
\multicolumn{2}{c}{RankEF-Soft}                         & 20.12          & 28.59          & 38.79          & 16.14          & 23.86          & 38.07          \\ \hline
                                        & $r$=1           & 17.73          & 26.25          & 33.95          & 13.38          & 21.37          & 32.69          \\
                                        & $r$=2           & 18.23          & 27.41          & 36.79          & 13.24          & 21.79          & 36.28          \\
                                        & \cellcolor[HTML]{DDDDDD}$r$=3 & \cellcolor[HTML]{DDDDDD}19.76 & \cellcolor[HTML]{DDDDDD}28.03 & \cellcolor[HTML]{DDDDDD}37.95 & \cellcolor[HTML]{DDDDDD}15.90 & \cellcolor[HTML]{DDDDDD}23.86 & \cellcolor[HTML]{DDDDDD}37.10 \\
\multirow{-4}{*}{RankEF-INF}            & $r$=4           & 17.89          & 26.58          & 35.12          & 13.97          & 21.97          & 35.76          \\ \hline
\end{tabular}
}
% \vspace{-1em}

%We use the fine-tuned CodeGen-2B as the base generation model.}
\label{tab:tabel_6}
\vspace{-1em}
\end{table}

\subsection{RQ4: Multi-Task Training Effectiveness}
\label{section: RQ4}
We incorporate three diverse multi-task training paradigms, namely, Hard Parameter Sharing (denoted as "Hard"), Soft Parameter Sharing (denoted as "Soft"), and Intermediate Fine-Tuning abbreviated as "INF"). For RankEF-INF, we conduct multiple rounds of experiments, where 1,000 steps of training on performing the feedback generation task and 1,000 steps of training on the classification task are considered as one round. Table~\ref{tab:tabel_6} shows that RankEF with Hard Parameter Sharing outperformed on both validation and test sets. Specifically, on the APPS test set, RankEF-Hard achieves 17.50\% Pass@1, surpassing RankEF-Soft's 16.14\% and RankEF-INF's 15.90\%.

Compared to RankEF-Soft, RankEF-Hard concurrently trains on two tasks within the same model. In the confines of our experimental setup, both the classification and generation tasks exhibit marked similarities. The inputs for these tasks are predominantly analogous, with only the initial identifiers differentiating them. They share parameters of the encoder component, wherein the function of the encoder can be conceptualized as a comprehension of both tasks. In our multi-task framework, enhancing the comprehension of both tasks concurrently is beneficial for augmenting the model's overall performance. Training both tasks within a unified model optimally facilitates the model's capacity to assimilate beneficial shared information.

RankEF-INF undergoes an initial training phase for the generation task, followed by a subsequent training phase for the classification task. Given the elongated interval between the two tasks, it is plausible that the model might exhibit partial amnesia regarding the nuances of the prior task, resulting in the sub-optimal performance of RankEF-INF. We try several rounds of experiments on this and find that RankEF-INF works best when $r = 3$, but is still inferior to RankEF-Hard. However, it is salient to note that regardless of the specific multi-task training paradigm employed, there is a notable performance enhancement in comparison to singular task training. This accentuates that the integration of execution feedback aids the model in discerning between correct and incorrect code.

% \begin{tcolorbox}[size=title]
% \textbf{Answer to RQ4:} \textcolor{black}{We explored three distinct multi-task training methods and found that all were effective for RankEF. Among them, RankEF-Hard demonstrated the best performance.}
% \end{tcolorbox}

\find{
\textbf{\ding{45} Answer to RQ4:} \textcolor{black}{We explored three distinct multi-task training methods and found that all were effective for RankEF. Among them, RankEF-Hard demonstrated the best performance.}
}

\subsection{RQ5: Task Weighting Parameters}
\label{section: RQ5}
\begin{table}[t]
\caption{Results of ablation experiments with different parameter weights on APPS. CodeGen (2B), after fine-tuning, serves as the base generation model.}
\vspace{-1em}
\resizebox{0.48\textwidth}{!} {
\begin{tabular}{ccccccc}
\hline
Task's Weight & \multicolumn{3}{c}{Val}                          & \multicolumn{3}{c}{Test~(Solved Problems)}        \\ \hline
$\lambda$        & Pass@1         & Pass@2         & Pass@5         & Pass@1         & Pass@2         & Pass@5         \\ \hline
0~(CodeRanker) & 19.23          & 27.92          & 37.29          & 13.96          & 20.69          & 33.93          \\ \hline
0.1           & 17.89          & 24.25          & 34.95          & 14.62          & 20.82          & 36.00          \\
0.3           & 19.40          & 25.59          & 35.28          & 16.28          & 23.03          & 38.43          \\
0.5           & 18.23          & 25.75          & 36.96          & 15.59          & 24.00          & 38.62          \\
0.7           & 21.40          & 27.42          & 38.79          & 17.03               & 26.20          & 40.66          \\
0.9           & \textbf{25.08} & \textbf{31.61} & \textbf{41.14} & \textbf{17.50} & \textbf{26.62} & \textbf{41.93} \\ \hline
1             & 17.77               & 24.02               & 33.29               &  13.25              & 20.42               & 34.01               \\ \hline
\end{tabular}
}
\vspace{-1em}
%We use the fine-tuned CodeGen-2B as the base generation model.}
\label{tab:tabel_10}
\end{table}
As shown in Table~\ref{tab:tabel_10}, we evaluate the performance of RankEF-Hard under various parameter weight settings. The results show that regardless of how the parameter weights are adjusted, the performance of RankEF-Hard is superior to that of single-task methods on test set. This finding demonstrates the robustness of RankEF in multitask learning; compared to single-task methods, it achieves better performance. Additionally, we note that RankEF's performance improves with increased emphasis on the execution feedback prediction task. This may be because this task's complexity encourages deeper, more comprehensive feature learning, boosting the model's overall capacity. Hence, emphasizing the execution feedback prediction task could enhance the model's utilization of these features, leading to better performance. 

However, it is worth noting that when we set $\lambda$ to 1, meaning that we rely solely on the execution feedback generation task to rank the candidate code, RankEF generates execution feedback to rank the code, rather than relying on an untrained MLP classifier. We found that its performance was not ideal. This is because the generation task itself is not particularly well-suited for ranking tasks. However, as an auxiliary task, it can further enhance RankEF's ranking ability, indirectly demonstrating the effectiveness of our multi-task approach.

% \begin{tcolorbox}[size=title]
% \textbf{Answer to RQ5:} \textcolor{black}{We experimented with varying task weights and observed that the performance of RankEF improves as the weight of the execution feedback generation task increases. However, when relying solely on the execution feedback generation task, RankEF's performance decreases substantially, indicating that both tasks are indispensable.}
% \end{tcolorbox}

\find{
\textbf{\ding{45} Answer to RQ5:} \textcolor{black}{We experimented with varying task weights and observed that the performance of RankEF improves as the weight of the execution feedback generation task increases. However, when relying solely on the execution feedback generation task, RankEF's performance decreases substantially, indicating that both tasks are indispensable.}
}

\section{Discussion}

% \subsection{How is RankEF's generalized ability?}

\begin{figure}[htbp]
    \centering
    \includegraphics[width=0.86\linewidth]{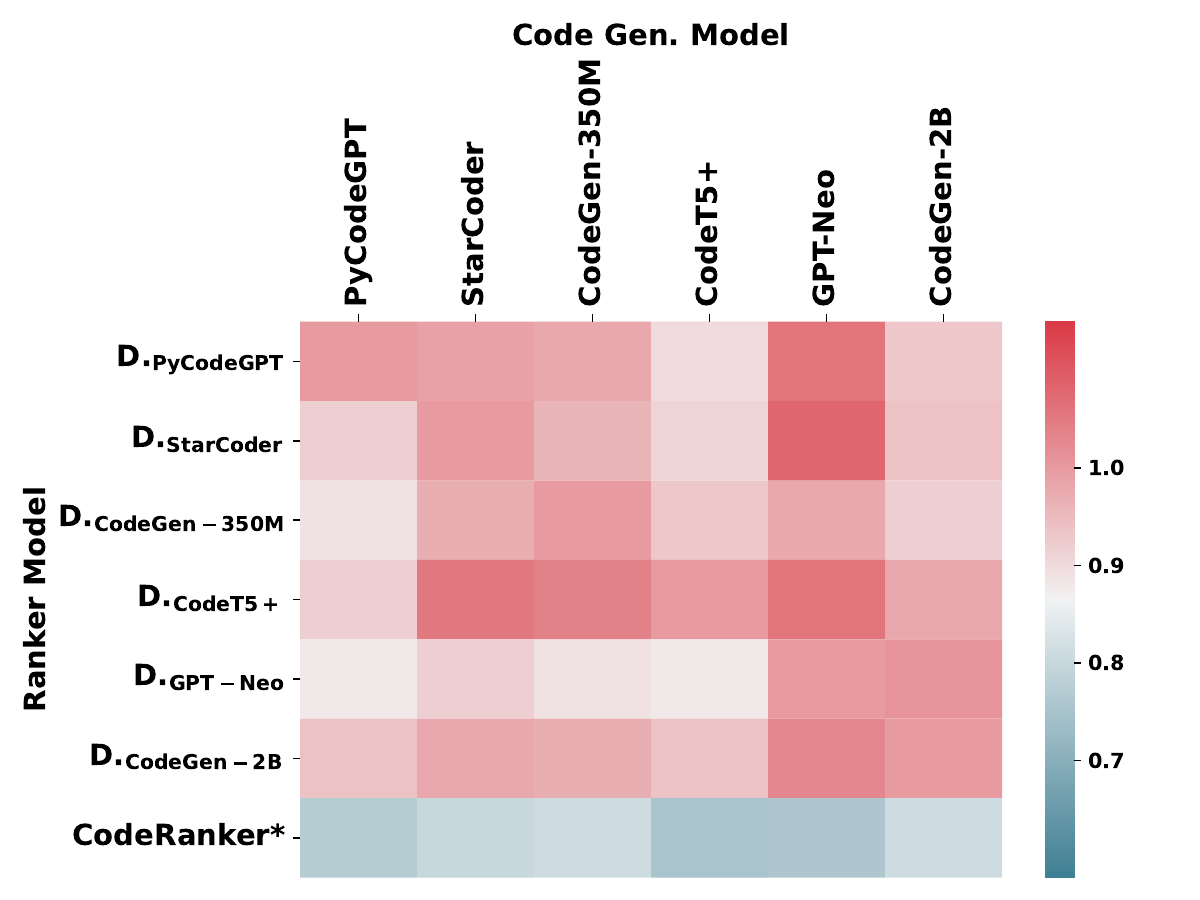}
    \vspace{-3mm}
    \caption{Results of RankEF's generalized capabilities on datasets with different models~(CodeRanker* is trained on the same model data).}
    \Description{Results of RankEF's generalized capabilities on datasets with different models~(CodeRanker* is trained on the same model data).}
    \label{fig: hot}
    \vspace{-3mm}
\end{figure}

\textbf{How is RankEF's generalized ability?}. 
% The original CodeRanker study highlights that rankers, when trained with data from a consistent code generation model, perform optimally. However, a significant performance dip occurs when using data from differing models, indicating a potential limitation in CodeRanker's generalizability. To delve deeper, we examined RankEF's generalization across various datasets, as illustrated in Figure~\ref{fig: hot}. We normalize the Pass@1 outcomes for diverse RankEF implementations on the APPS test set, benchmarking RankEF utilizing consistent model data to a score of 1.
The original CodeRanker study shows optimal ranker performance with consistent code generation model data but reveals a drop when using data from different models, indicating limited generalizability. We examined RankEF's generalization across datasets (see Figure~\ref{fig: hot}), normalizing Pass@1 results on the APPS test set to a score of 1 for RankEF with consistent model data.

\begin{figure*}
    \centering
    \begin{subfigure}[b]{0.49\linewidth}
        \centering
        \includegraphics[width=\linewidth]{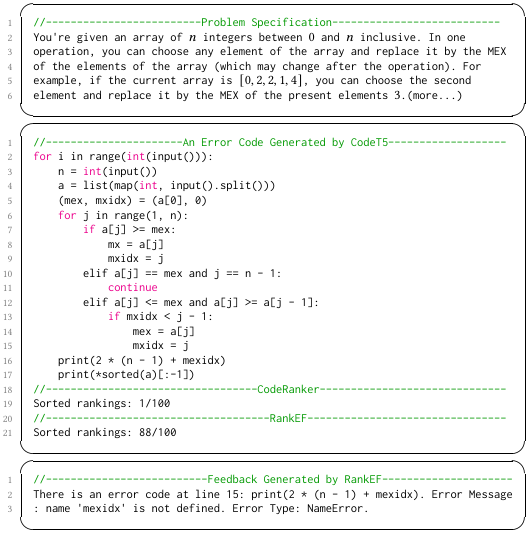}
        \caption{Successful Ranking.}
        \label{fig:case_1}
    \end{subfigure}
    \hfill
    \begin{subfigure}[b]{0.49\linewidth}
        \centering
        \includegraphics[width=\linewidth]{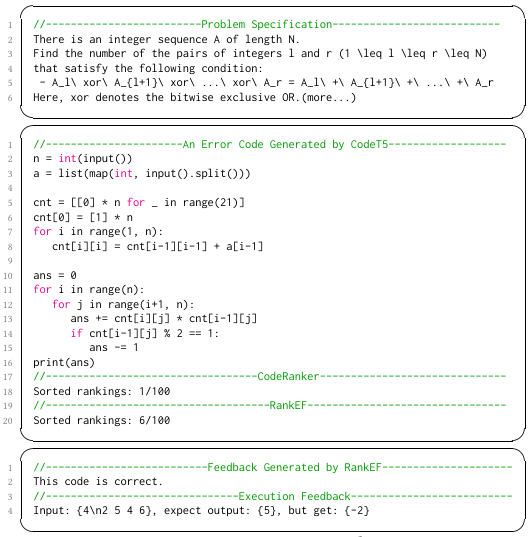}
        \caption{Failed Ranking.}
        \label{fig:case_2}
    \end{subfigure}
    \vspace{-1em}
    \caption{Two examples of RankEF ranking successes and failures respectively.}
    \Description{Two examples of RankEF ranking successes and failures respectively.}
    \label{fig:case}
    \vspace{-1em}
\end{figure*}

Our findings reveal that RankEF consistently delivers commendable performance across models, regardless of the code generation model’s data source. Notably, RankEF excels when trained on CodeT5+ data. Additionally, RankEF models trained on varied datasets consistently outperform CodeRanker versions trained on a single model’s data. This superior performance is due to RankEF leveraging execution feedback, enabling it to genuinely discern the essence of code errors. In contrast, CodeRanker, which relies solely on error type labels, is susceptible to biases in dataset-specific error distributions, limiting its generalizability.

% \subsection{Why not use LLMs (e.g. ChatGPT) for code ranking?}

\noindent \textbf{Why not use LLMs (e.g., ChatGPT) for code ranking?} In the field of natural language processing, recent research~\cite{sun-etal-2023-chatgpt, wang-etal-2023-chatgpt, Dai-chatgpt} has focused on utilizing LLMs, such as ChatGPT, for text ranking tasks. We attempted to use ChatGPT (GPT-3.5-Turbo) with the following prompt to rank code: "Please determine if the code below matches the problem description:". However, we found that GPT-3.5-Turbo could not accurately distinguish between correct and incorrect code for complex programming problems. Furthermore, we believe that using ChatGPT to rank a large number of code candidates is not a prudent choice, as the cost of invoking ChatGPT's API is based on the number of tokens, making the use of ChatGPT expensive when dealing with a large volume of code candidates.

\section{Qualitative Analysis}

% In this section, we aim to investigate whether RankEF can truly understand the root causes of erroneous code. Figure ~\ref{fig: case_study} shows a candidate erroneous code generated by CodeT5 based on the problem description. When this code is executed, the compiler returns an error message, indicating that the code in line 15, "print(2 * (n - 1) + mexidx)," uses an undefined variable name "mexidx." We asked RankEF to analyze the cause of this error based on the problem description and the code. The results show that RankEF can correctly identify the cause of the erroneous code, demonstrating that our method effectively enables RankEF to understand the root cause of the code errors.

While statistical metrics provide valuable insights, they may not fully reflect whether RankEF truly understands why errors occur in the code. Therefore, we qualitatively analyzed examples of RankEF successes and failures separately as illustrated in Figure ~\ref{fig:case}.

Figure ~\ref{fig:case_1} presents an example where CodeRanker fails to rank correctly while RankEF succeeds. This is an erroneous code snippet generated by CodeT5 from a set of 100 codes. CodeRanker placed this code in the first position, whereas RankEF ranked it 88th. Due to RankEF's multi-task training approach, we further asked RankEF to analyze the cause of the error in this code. It can be seen that RankEF accurately identified the error: the variable "mexidx" on line 15 was not defined earlier in the code, leading to a NameError. However, CodeRanker failed to detect this subtle error and mistakenly judged the code to be correct. This phenomenon indicates that CodeRanker, which relies solely on code execution results as classification labels for training, cannot accurately understand the causes of errors. It merely judges based on superficial code features. In contrast, RankEF considers more useful information from execution feedback during the training process, which helps RankEF to more comprehensively understand the causes of code errors.

Figure ~\ref{fig:case_2} presents an example where both CodeRanker and RankEF fail to rank correctly. CodeRanker places this code in the first position, while RankEF ranks it sixth, both relatively high positions. We found that this code contains an intent error, meaning there are no syntax or execution errors, but the actual output does not match the expected output. Such errors are particularly challenging for rankers as they require a deep understanding of whether the code's logic fulfills the problem description. Although RankEF ranks it relatively high, correct code can still be found before this position. This indicates that while both CodeRanker and RankEF struggle with recognizing this type of error, RankEF's consideration of input-output feedback during training allows it to better handle such errors compared to CodeRanker.
\vspace{-2mm}

\section{Threats to Validity}
We have identified the following threats to validity:

\noindent\textbf{The nature of code ranking}. The essence of code ranking lies in selecting the correct code from a large pool of candidate codes. Although code ranking does not inherently enhance the model's performance, it serves as a post-processing technique that can significantly boost the model's capabilities. In the future, with the advent of extremely powerful models, code ranking may become unnecessary. However, given the current models' subpar performance when dealing with complex problems, code ranking remains a valuable post-processing method. Moreover, code ranking can complement other methods that inherently improve model performance, resulting in orthogonal enhancement effects.

\noindent \textbf{Overfitting on APPS}. To ensure a fair comparison, we follow the same settings as in the original CodeRanker paper~\cite{inala2022fault} (using code generation models fine-tuned on APPS) when evaluating RankEF's transferability across other datasets. However, this approach poses the threat of overfitting to the APPS style, especially in datasets like HumanEval, where the task format differs from APPS and MBPP. While HumanEval involves code completion based on partial code, APPS and MBPP require generating complete code based on natural language descriptions. The use of models fine-tuned on APPS for HumanEval may result in generating entire code from scratch rather than completing existing code. To mitigate this threat, in addition to using code generation models fine-tuned on APPS, we also select several larger base models for the  transferability experiments.

\noindent\textbf{Selection of dataset}. In this study, we select only one language, specifically the Python dataset, because Python datasets are the most popular in the field of code generation. Therefore, we have not claimed that our method is applicable to all programming languages. However, the core idea of RankEF is universal and can be applied to any programming language. In future work, we will evaluate the performance of RankEF on datasets of other programming languages to further validate its broad applicability.

\section{Related Work}

\textbf{Code Ranking for Code Generation.} 
With the advancement of artificial intelligence, numerous LLMs have emerged. Closed-source models like AlphaCode~\cite{li2022competition} and ChatGPT~\cite{ChatGPT}, along with open-source models such as CodeT5~\cite{wang2021codet5}, InCoder~\cite{fried2022incoder}, StarCoder~\cite{li2023starcoder}, Code Llama~\cite{codellama}, and WizardCoder~\cite{luo2023wizardcoder}, have demonstrated significant capabilities in certain software engineering tasks~\cite{ryan2024code, li2024ircoco, nam2024using}, especially in code generation~\cite{gu2023llm, sun2024enhancing,huang2024knowledge, jiang2024survey}. However, despite their impressive performance in code generation, these models often struggle with complex programming problems. Observing that utilizing LLMs to generate numerous candidate solutions increases the likelihood of producing correct code, researchers are increasingly focusing on selecting the correct code from these candidates, known as code ranking.

Current endeavors in code ranking can be categorized into two groups: execution-based~\cite{chen2022codet, shi2022natural, li2022competition, MBR-EXEC} and non-execution-based~\cite{inala2022fault, zhang2023coder}. These execution-based approaches require code execution on unit tests, which is not desirable in practice due to the insecurity of the model-generated code and the lack of unit tests. To address this limitation, CodeRanker~\cite{inala2022fault} trained a ranker using code executions as classification labels. Coder-Reviewer~\cite{zhang2023coder} ranked candidates by calculating the likelihood of the model's input code backpropagating the problem description. The former uses only classification labels and does not enable the ranker to understand the nature of the code errors, and the latter uses the code to invert the problem description and does not apply to complex problem descriptions. Another line of work~\cite{ni2023lever} trains a verifier based on natural language descriptions, the program itself, and its execution results to determine code correctness. This approach integrates generation and verification probabilities to re-rank candidate programs. Although it uses a trained verifier for ranking, it relies on execution results during both training and ranking phases. Thus, it is an execution-based code ranking method, sharing the same potential limitations as other execution-based methods.

\noindent\textbf{Code Generation with Feedback.} Recently, significant research progress has been made on enhancing model capabilities using feedback. For example, Self-debug~\cite{chen2023teaching}, Self-refine~\cite{madaan2024self}, and Reflexion~\cite{shinn2024reflexion} utilize powerful closed-source LLMs to iteratively generate code based on feedback obtained from the model itself or externally. Although these methods have demonstrated excellent performance, they rely on closed-source LLMs, and the process of iterative code generation leads to higher inference costs. 

Additionally, Self-edit~\cite{zhang-etal-2023-self} modifies code generated by LLMs by training an independent code editor. This method involves executing the code to obtain execution feedback, which is then used to train the code editor. The editor takes the code to be edited along with the execution feedback as input and outputs the edited code. While the code editor can edit code based on execution feedback, it also requires execution feedback during the inference phase, which is impractical in environments lacking unit tests or where the code is non-executable. In contrast, our approach trains an independent code ranker using a multi-task learning framework, eliminating the dependence on execution feedback during the inference phase, thereby overcoming the limitations of the aforementioned methods. 

% \noindent\textbf{Multi-Task Learning.} 
% % \noindent\textbf{Multi-Task Learning.}
% Multi-task learning, which leverages synergistic learning across multiple tasks, has proven instrumental in enhancing model performance on the individual task in natural language processing~\cite{radford2019language, oshingbesan2022extreme}. Recently, this paradigm has been extended to the realm of code, specifically in code completion~\cite{liu2020multi, izadi2022codefill, takerngsaksiri2022syntax}. These works combine code-completion task with code-type prediction to enhance code-completion. Other code domains, such as code summarization~\cite{wei2019code, wang2021mulcode, xie2021exploiting} and code generation~\cite{phan2021cotext, yang2023syntax}, multi-task learning has also exhibited commendable performance. In this paper, we pioneer multi-task learning for code ranking.

\section{Conclusion}
In this paper, we have proposed a novel approach termed RankEF, which harnesses execution feedback to enhance the efficiency of code ranking. 
Through the integration of execution feedback and classification labels using multi-task learning, RankEF not only distinguishes between accurate and erroneous code but also comprehensively understands the underlying factors contributing to diverse code errors. Experimental results show that, owing to its profound grasp of error causality, RankEF demonstrates superior performance in comparison to established baseline methods.

%\section{Data Availability}
\noindentparagraph{\textbf{\textup{Data Availability.}}}
% \noindent\textbf{Data Availability.}
All the experimental dataset and source code are available at: \url{https://github.com/sssszh/RankEF}.

%%
%% The acknowledgments section is defined using the "acks" environment
%% (and NOT an unnumbered section). This ensures the proper
%% identification of the section in the article metadata, and the
%% consistent spelling of the heading.
\begin{acks}
The work is supported in part by the Natural Science Foundation of Shandong Province, China (Grant No. ZR2021MF059), the National Natural Science Foundation of China (Grant Nos. 62192731, 62072007, 62192733, 61832009, 62192730), the National Key R\&D Program (Grant No. 2023YFB4503801) and the Major Program (JD)
of Hubei Province (Grant No. 2023BAA024)).
\end{acks}

%%
%% The next two lines define the bibliography style to be used, and
%% the bibliography file.
\bibliographystyle{ACM-Reference-Format}
% \bibliographystyle{unsrt}
% \bibliography{sample-base}

%%
%% If your work has an appendix, this is the place to put it.

\end{document}